\documentclass[a4paper,fleqn]{cas-dc}

\usepackage[numbers]{natbib}

\def\tsc#1{\csdef{#1}{\textsc{\lowercase{#1}}\xspace}}
\tsc{WGM}
\tsc{QE}
\tsc{EP}
\tsc{PMS}
\tsc{BEC}
\tsc{DE}
\begin{document}
\let\WriteBookmarks\relax
\def\floatpagepagefraction{1}
\def\textpagefraction{.001}
\shorttitle{CellResDM}
\shortauthors{X. Xing et~al.}

\title [mode = title]{Artificial Immunofluorescence in a Flash: Rapid Synthetic Imaging from Brightfield Through Residual Diffusion}                      
\tnotetext[1]{This study was supported in part by the ERC IMI (101005122), the H2020 (952172), the MRC (MC$\slash$PC$\slash$21013), the Royal Society (IEC$\backslash$NSFC$\backslash$211235), the NVIDIA Academic Hardware Grant Program, the SABER project supported by Boehringer Ingelheim Ltd, NIHR Imperial Biomedical Research Centre (RDA01), Wellcome Leap Dynamic Resilience, UKRI guarantee funding for Horizon Europe MSCA Postdoctoral Fellowships (EP$\slash$Z002206$\slash$1), and the UKRI Future Leaders Fellowship (MR$\slash$V023799$\slash$1).}

% \tnotetext[2]{The second title footnote which is a longer text matter
%    to fill through the whole text width and overflow into
%    another line in the footnotes area of the first page.}

%%%%%author information
%%%%%author information
\author[1]{Xiaodan Xing}
% \cormark[1]
\fnmark[1]
% \ead{xxing@imperial.ac.uk}
% \credit{Conceptualization, Methodology, Software}
\affiliation[1]{organization={Bioengineering Department and Imperial-X, Imperial College London},
                city={London},
                % postcode={SW7 2AZ}, 
                country={United Kingdom}}

\author[2]{Chunling Tang}
\fnmark[1]
\affiliation[2]{organization={Centre for Craniofacial \& Regenerative Biology, King’s College London},
                city={London},
                % postcode={SE1 9RT}, 
                country={United Kingdom}}
\author[1]{Siofra Murdoch} 
% \fnmark[1]
\author[3,4]{Giorgos Papanastasiou}
\affiliation[3]{organization={Archimedes Unit, Athena Research Centre},
                city={Athens},
                % postcode={EH16 4TJ}, 
                country={Greece}}
\affiliation[4]{organization={School of Computer Science and Electronic Engineering, The University of Essex},
                city={Essex},
                % postcode={CO4 3SQ}, 
                country={United Kingdom}}                
% \author[5]{Jan Cross-Zamirski}
% \affiliation[5]{organization={Department of Applied Mathematics and Theoretical Physics, University of Cambridge},
%                 city={Cambridge},
%                 % postcode={CB3 0WA}, 
%                 country={United Kingdom}} 

\author[2]{Yunzhe Guo}

\author[1]{Xianglu Xiao}

\author[5]{Jan Cross-Zamirski}
\affiliation[5]{organization={Department of Applied Mathematics and Theoretical Physics, University of Cambridge},
                city={Cambridge},
                % postcode={CB3 0WA}, 
                country={United Kingdom}} 

\author[5]{Carola-Bibiane Schönlieb}

\author[6]{Kristina Xiao Liang}
\affiliation[6]{organization={Department of Clinical Medicine (K1), University of Bergen},
                city={Bergen},
                % postcode={CB3 0WA}, 
                country={Norway}}

\author[7]{Zhangming Niu}
\affiliation[7]{organization={MindRank AI Ltd.},
                city={Hangzhou},
                % postcode={CB3 0WA}, 
                country={China}}

\author[8]{Evandro Fei Fang}
\affiliation[8]{organization={Department of Clinical Molecular Biology, University of Oslo and Akershus University Hospital,},
                city={Lørenskog},
                % postcode={CB3 0WA}, 
                country={Norway}}

\author[9]{Yinhai Wang}
\affiliation[9]{organization={Data Sciences and Quantitative Biology, Discovery Sciences, AstraZeneca R\&D},
                city={Cambridge},
                % postcode={CB3 0WA}, 
                country={United Kingdom}}

\author[1,10,11,12]{Guang Yang}
\affiliation[10]{organization={National Heart and Lung Institute, Imperial College London},
                city={London},
                % postcode={CB3 0WA}, 
                country={United Kingdom}} 
\affiliation[11]{organization={School of Biomedical Engineering \& Imaging Sciences, King's College London},
                city={London},
                % postcode={CB3 0WA}, 
                country={United Kingdom}} 
\affiliation[12]{organization={Cardiovascular Research Centre, Royal Brompton Hospital},
                city={London},
                % postcode={CB3 0WA}, 
                country={United Kingdom}} 

\cormark[1]
\fntext[fn1]{Xiaodan and Chunling contributed equally to this work.}
\ead{gyang@imperial.ac.uk}

\cortext[cor1]{Corresponding author}
% \fntext[fn1]{Xiaodan, Siofra and Chunling contributed equally to this work.}

\begin{abstract}
Immunofluorescent (IF) imaging is crucial for visualizing biomarker expressions, cell morphology and assessing the effects of drug treatments on sub-cellular components. IF imaging needs extra staining process and often requiring cell fixation, therefore it may also introduce artefects and alter endogenouous cell morphology. Some IF stains are expensive or not readily available hence hindering experiments. Recent diffusion models, which synthesise high-fidelity IF images from easy-to-acquire brightfield (BF) images, offer a promising solution but are hindered by training instability and slow inference times due to the noise diffusion process. This paper presents a novel method for the conditional synthesis of IF images directly from BF images along with cell segmentation masks. Our approach employs a Residual Diffusion process that enhances stability and significantly reduces inference time. We performed a critical evaluation against other image-to-image synthesis models, including UNets, GANs, and advanced diffusion models. Our model demonstrates significant improvements in image quality ($p<0.05$ in MSE, PSNR, and SSIM), inference speed (26 times faster than competing diffusion models), and accurate segmentation results for both nuclei and cell bodies (0.77 and 0.63 mean IOU for nuclei and cell true positives, respectively). This paper is a substantial advancement in the field, providing robust and efficient tools for cell image analysis.
\end{abstract}

\begin{graphicalabstract}
\includegraphics[width=1\linewidth]{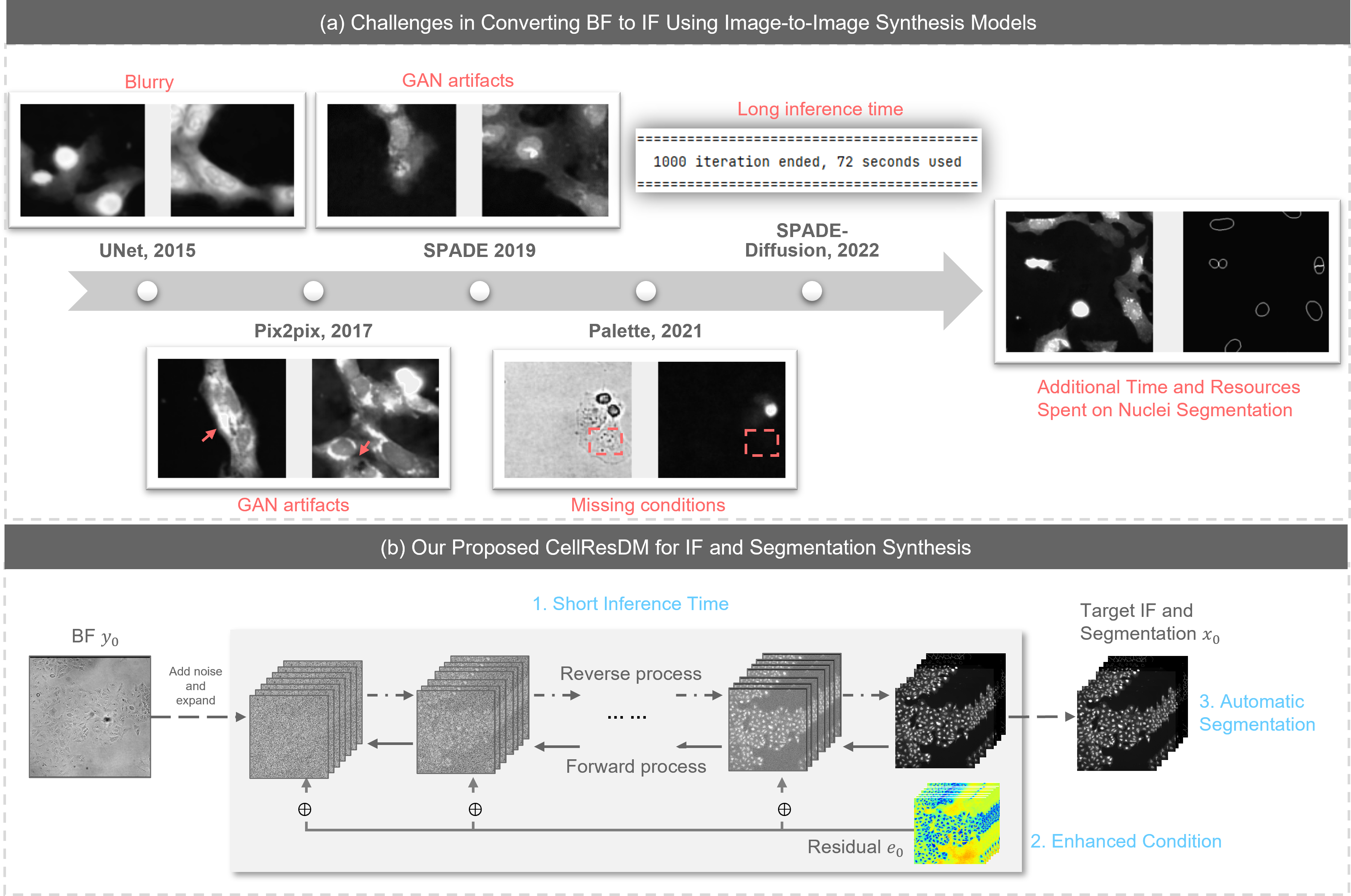}
\end{graphicalabstract}

\begin{highlights}
\item We introduce CellResDM, a novel residual diffusion model to synthesise immunofluorescence (IF) images from brightfield (BF) images with enhanced stability and performance.
\item Our CellResDM approach improves synthesised image quality, inference speed, and segmentation accuracy for nuclei and cell body, outperforming existing image-to-image synthesis models.
\item CellResDM model simultaneously generates synthetic IF images and cell/nuclei segmentation, streamlining the process.
\end{highlights}


\maketitle

\section{Introduction}
The cell painting technique \cite{bray2016cell} employs a suite of fluorescent reagents to visualise and analyse the spatial organization of subcellular structures and components. It is increasingly used in drug screening to evaluate the effects of small molecules or other drug modalities on cells. Indeed, the patterns of cellular staining can reveal morphological differences in treated samples. Through cell painting, researchers can investigate the dynamic organization of proteins, which further explains the functions states of cells such as cell viability and proliferation, as well as drug mechanism of actions (MoAs) such as efficacy, toxicity and DNA damage.

Traditional methods of evaluating immunofluorescence (IF) images often face limitations due to the restricted number of imaging channels available (4-5). IF staining lab processes may introduce extra artefacts which are either related to or directly causes batch effects between samples, which makes the comparison of data across wells/plates very difficult.

%Recent study ([1], In Silico Labeling: Predicting Fluorescent Labels in Unlabeled Images, [2] Jan et al, 2023, etc.) showed the possiblity of predicting IF channels from brightfield which suggests that brightfield images do contain hidden IF biomarker information, however without IF artefacts, therefore we in this study would like to continue from these efforts to make the prediction of IF stains more practical and accurate. 

Recent advances in generative deep learning models \cite{christiansen2018silico,cross2022label}, showed the possibility of predicting IF channels from brightfield (BF) which suggests that BF images do contain hidden IF biomarker information, however without IF artefacts. For example, CNN-based models \cite{christiansen2018silico} and generative adversarial networks (GANs) \cite{lee2021deephcs++,cross2022label} have been proposed to efficiently synthesise IF images from BF images.

\begin{figure*}
    \centering
    \includegraphics[width=1\linewidth]{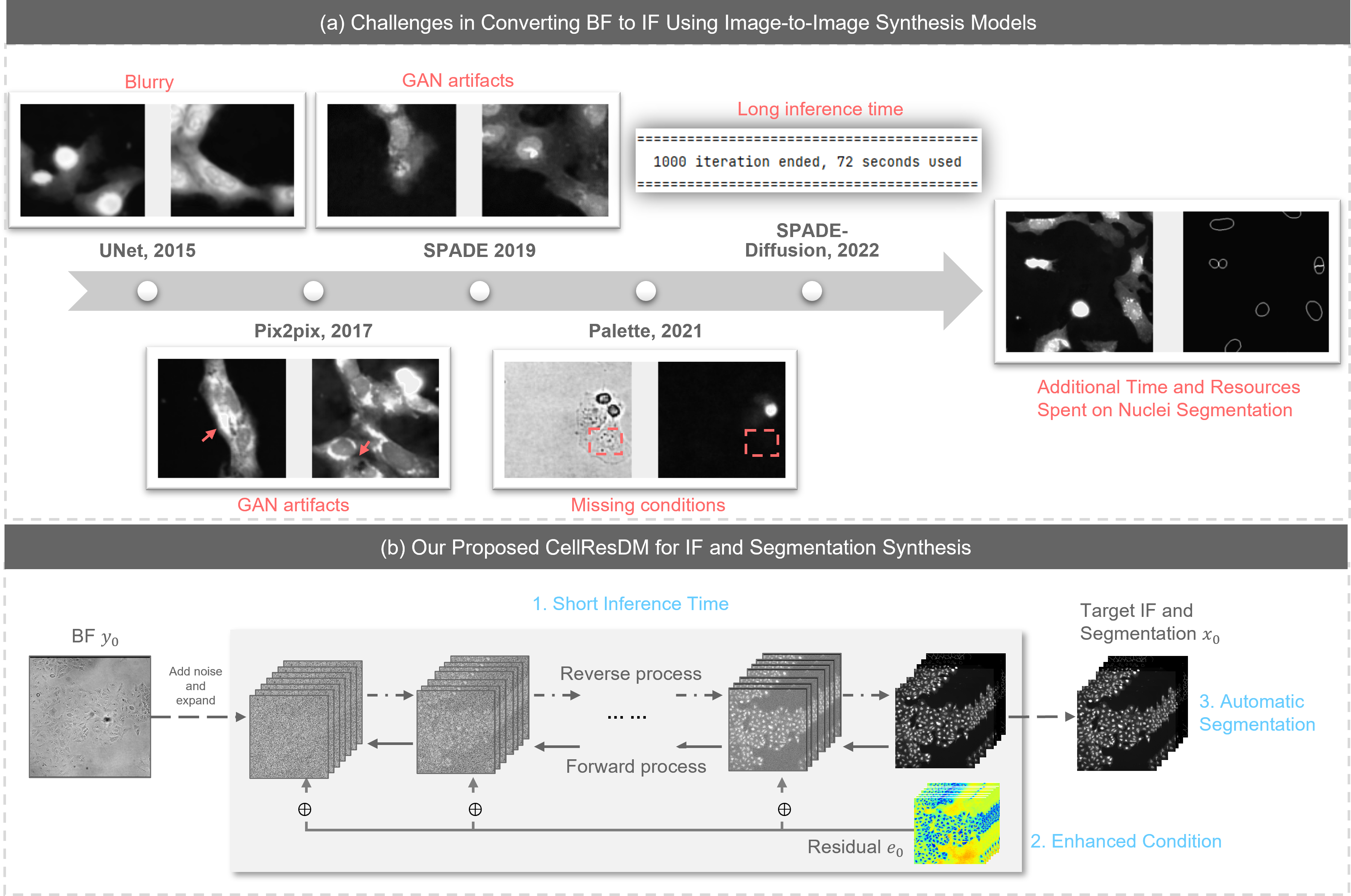}
    \caption{The limitations of current image-to-image synthesis models on the BF to IF synthesizing task (a) and our proposed CellResDM for IF and Segmentation Synthesis (b). Our proposed CellResDM can reduce the inference time for virtual cell painting and produce accurate nuclei and cell segmentations simultaneously.}
    \label{fig:teaser}
\end{figure*}

However, CNN-based models can produce blurred results, and GAN-based models face challenges in creating images without hallucinations (Fig. \ref{fig:teaser} (a)). Diffusion models have been proposed to address these limitations, but they encounter significant challenges in training stability and prolonged inference times. This is because diffusion models do not directly optimise the final image, but, instead, they optimise through each step of the diffusion process. Moreover, these models often face challenges in maintaining consistent performance across different datasets, a critical requirement for reliable drug discovery processes. Additionally, obtaining the synthesised IF images is not the final aim of the drug screening. Single-cell analysis also requires accurate nuclei and cell segmentation, which can be time-consuming.

To address these shortcomings, this paper introduces a novel application of residual diffusion models \cite{yue2024resshift} tailored for the conditional synthesis of IF staining images directly from BF images and accompanying cell segmentation masks, as in Fig. \ref{fig:teaser} (b). Our approach aims to streamline the generative and segmentation process using a Residual Diffusion process, improving the stability of the IF prediction performance and reducing the inference time required by standard diffusion. Further, we compare our methods against existing generative models that have been developed for cell painting IF synthesis. Our models demonstrate significant improvements in terms of image quality of synthetic images, inference speed and segmentation accuracy of nuclei and other cell regions. 

Our contributions are threefold. First, we evaluate and summarise the shortcomings of current deep learning-based methods for BF to IF image synthesis. Second, we develop a novel residual diffusion model for the synthesis of IF images from BF images, enhancing the stability and performance of IF image prediction. Third, we propose the simultaneous accurate segmentation of nuclei and other cell regions, significantly enhancing the analysis of cell images by reducing the time required for downstream analysis. 

These innovations provide a robust framework that not only improves the efficiency of generating high-quality IF images but also integrates segmentation directly within the synthesis process, addressing critical needs in biological research and pharmaceutical testing. Compared to conventional segmentation \cite{stirling2021cellprofiler} and deep learning based segmentation algorithms \cite{stardist,stringer2021cellpose}, our method promise fast and accurate segmentation without additional training. The performance improvements highlighted by our comparative studies emphasize the potential of our method to set new benchmarks in the field of image synthesis and automated cellular analysis. This work paves the way for more accurate and timely insights into cellular behaviors, ultimately enhancing drug discovery and biology research.

\section{Related Works}
\subsection{BF to IF Image Synthesis}

In deep learning, predicting IF images from BF images is treated as an image translation task. Table \ref{table:ai_algorithms} summarises some of the recent algorithms currently used in IF image synthesis and their specific target IF channels.

Christiansen et al. \cite{christiansen2018silico} utilised convolutional neural networks (CNNs) optimized with Mean Squared Error (MSE) loss to match predicted with actual IF images. This method, however, often resulted in blurred details due to the MSE loss. Lee et al. \cite{lee2018deephcs} improved on this by incorporating a Transformation Network for initial image conversion and a Refinement Network for enhancing details. The advent of Generative Adversarial Networks (GANs) in 2014 marked a significant advance in image synthesis, outperforming traditional CNNS by leveraging a generator to create images and a discriminator to assess them \cite{helgadottir2021extracting,lee2021deephcs++,cross2022label}. 

More recently, diffusion models have been developed, providing higher fidelity and diversity \cite{ho2020denoising} by sequentially adding and then removing noise to recreate images. Despite it is quite hard and time consuming to do training \cite{cross2023class}, diffusion models face considerable challenges in training and inference. Indeed, they require prolonged inference times due to their iterative denoising process. Furthermore, conditional diffusion models often struggle with stability, sometimes overlooking specified input conditions \cite{tai2023revisiting,zhang2024enhancing}, which can affect the consistency and reliability of the generated images.

\begin{table*}[h]
\centering
\begin{tabular}{m{5cm}m{1.5cm}m{2.5cm}m{7cm}}
\hline
\textbf{Paper Name} & \textbf{Year} & \textbf{Model} & \textbf{Output} \\ \hline
In Silico Labeling \cite{christiansen2018silico} & 2018 & CNN & Nuclei (Hoechst, DAPI), Neurons (anti-TuJ1), Dendrites (anti-MAP2), Dead Cells (Propidium Iodide), Cell Membrane (CellMask) \\ \hline
DeepHCS \cite{lee2018deephcs} & 2018 & CNN & Nuclei (DAPI) \\ \hline
Extracting Quantitative Biological Information \cite{helgadottir2021extracting} & 2020 & GAN & Nuclei (Hoechst), Cell Cytoplasm (CellTracker Deep Red), Lipid Droplets (Bodipy) \\ \hline
DeepHCS++ \cite{lee2021deephcs++} & 2021 & GAN & Dead Cells (Alexa 488), Nuclei (Hoechst 33342), Cell Shape (Alexa 594) \\ \hline
Label-free Prediction \cite{cross2022label} & 2022 & WGAN & DNA (Hoechst 33342), Endoplasmic Reticulum (Concanavalin A), Nucleoli and Cytoplasmic RNA (SYTO 14), Actin (Phalloidin), Golgi and Plasma Membrane (WGA), Mitochondria (MitoTracker Deep Red) \\ \hline
Class-guided Image-to-image Diffusion \cite{cross2023class} & 2023 & Diffusion Model & DNA (Hoechst 33342), Endoplasmic Reticulum (Concanavalin A), Nucleoli and Cytoplasmic RNA (SYTO 14), Actin (Phalloidin), Golgi and Plasma Membrane (WGA), Mitochondria (MitoTracker Deep Red) \\ \hline
\end{tabular}
\caption{Summary of generative AI algorithms and their target IF channels for IF image synthesis from BF images.}
\label{table:ai_algorithms}
\end{table*}

\subsection{Diffusion Models}
Denoising Diffusion Probabilistic Models (DDPMs) \cite{ho2020denoising} have emerged as a powerful class of generative models capable of producing high-quality samples by modeling the data distribution through a diffusion process. This section provides an overview of the original DDPM framework, detailing both the forward and reverse processes, and laying the mathematical foundation for the proposed residual diffusion models. Furthermore, we provide theoretical explanations for causes of the missing conditions and the inference time issues associated with DDPM.

\textbf{Forward Process.} The forward process in DDPMs is designed to gradually add noise to a data sample, transforming it into Gaussian noise through a series of time steps. This process can be described by a Markov chain where the state at each time step $t$ is denoted as $x_t$. At each time step $t$, the data transforms as follows:
\begin{equation}
    x_t = \alpha_t x_{t-1} + \beta_t \epsilon_t,
\end{equation}

where $\alpha_t,\beta_t>0$, and $\alpha_t^2 + \beta_t^2=1$. Here, $\beta_t$ represents the noise variance schedule, and $\epsilon_t \sim \mathcal{N}(0, I)$ is Gaussian noise. The process starts from the original data $x_0$ and progresses to the final state $x_T$, which approximates a standard Gaussian distribution. The forward process ensures a smooth transition from data to noise, facilitating the training of the reverse process. 

\textbf{Reverse Process.} The reverse process in DDPMs aims to reconstruct the original data from the noised state by learning the reverse transitions. The target is to use a neural network parameterised by $\theta$ to estimate the mapping from $x_t$ to $x_{t-1}$. Let this model be $\mu(x_t)$, then a straightforward learning scheme is to minimise the Euclidean distance between the two states:
\begin{equation}
\mathcal{L}_t = \|x_{t-1} - \mu(x_t)\|^2
\end{equation}

Considering that 
\begin{equation}
x_{t-1} = \frac{1}{\alpha_t}(x_t - \beta_t \epsilon_t),
\end{equation}
to refine the loss function further, the estimation function is re-parameterised as 

\begin{equation}\label{eq:reparam}
\mu(x_t) = \frac{1}{\alpha_t}(x_t - \beta_t \epsilon_\theta(x_t, t)).
\end{equation}
Substituting this into the loss function, we get
\begin{equation}
\mathcal{L}_t = \|x_{t-1} - \mu(x_t)\|^2 = \frac{\beta_t^2}{\alpha_t^2} \|\epsilon_t - \epsilon_\theta(x_t, t)\|^2.
\end{equation}

The neural network $\epsilon_\theta(x_t, t)$ predicts the added noise, and the objective is to minimise the difference between the predicted noise and the actual noise. By minimising this objective, the model learns to effectively reverse the noise addition, enabling the generation of high-quality samples from noise. For image-to-image translation tasks where the input image condition is $y_0$, the optimisation function can thus be revised as 
\begin{equation}
\mathcal{L}_t = \frac{\beta_t^2}{\alpha_t^2} \|\epsilon_t - \epsilon_\theta(x_t, y_0,t)\|^2.
\end{equation}

Conditional DDPMs can be implemented using either classifier guidance or classifier-free guidance, where the latter is achieved through condition drop-outs. Although we will not discuss the conditioning techniques in detail in this section, it is evident that the original DDPM optimisation function does not directly optimise the relationship between $y_0$ and $x_0$. Furthermore, it is worth noting that the diffusion step $T$ is typically set to $10^3$, resulting in long inference time.

\subsection{Nuclei and Cell Segmentation}
Nuclei and cell segmentation are critical tasks in biomedical image analysis, enabling the quantitative study of cellular structures and their functions. Early approaches to nuclei and cell segmentation primarily relied on classical image processing techniques such as thresholding \cite{otsu1979threshold}, edge detection, and watershed algorithms \cite{roerdink2000watershed}. However, these traditional methods often proved inadequate for images with complex backgrounds, overlapping cells, or varying cell sizes and shapes. To address these limitations, Convolutional Neural Networks (CNNs) have since been widely applied in nuclei and cell segmentation tasks. Further, researchers have introduced deep learning algorithms based on CNN architectures, such as StarDist \cite{stardist}, Cellos \cite{mukashyaka2023cellos} and CellPose \cite{stringer2021cellpose}, that have improved performance. 

However, challenges remain in nuclei and cell segmentation. These models require additional GPU resources for training and require large annotated datasets. Xing et al. \cite{xing2024segmentanything} used foundation models \cite{kirillov2023segment} for 2D organoid detection using brightfield images, which can be applicable to nuclei and cell segmentation. However, this algorithm still requires post-processing and can be time-consuming for large scale image analysis.

Intuitively, a data synthesis model that can draw realistic images must contain rich semantic information \cite{zhang2021datasetgan}. Thus, we hypothesise that IF synthesis models should have the ability to produce accurate segmentation masks at the same time. As such, in this paper, we propose the adoption of a hybrid setting to synthesise images and produce the segmentation masks at the same time.

\section{Methods}
\subsection{Residual Diffusion}
\begin{figure}[h]
    \centering
    \includegraphics[width=1\linewidth]{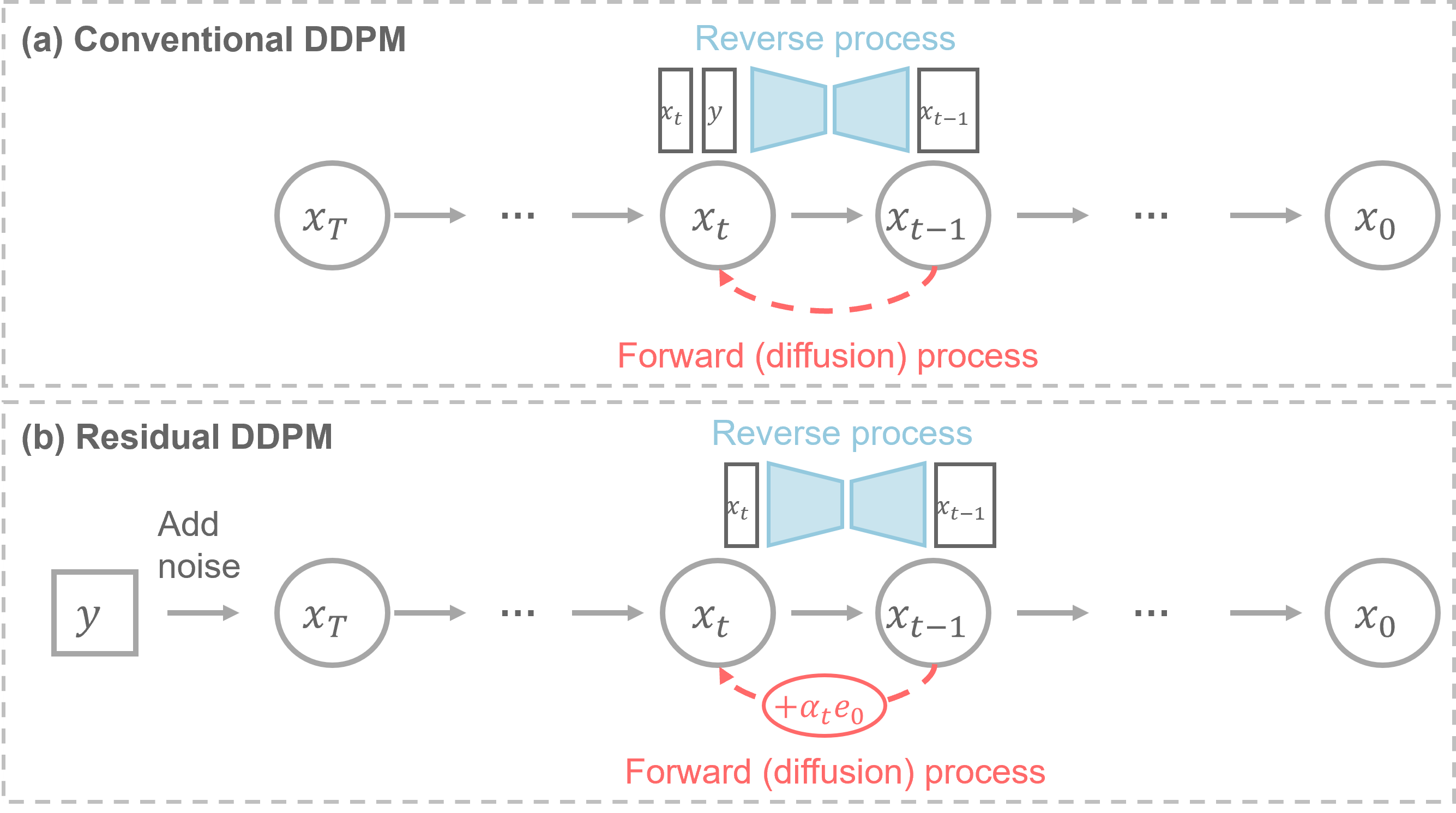}
    \caption{Overview of the original DDPM process (a) and the Residual Diffusion process (b) in this paper. The Residual Diffusion process progressively transforms the condition BF image into the target IF image by incrementally incorporating the residuals between the condition and the target images.}
    \label{fig:resshift}
\end{figure}

In the Related Work section, we discussed how traditional diffusion models compute the gradual transformation of data from a structured state into a random state and then learn to reverse this process to generate new data samples. Inspired by \cite{yue2024resshift}, who revised the diffusion process for super-resolution, we similarly applied the residual diffusion concept to the BF to IF image translation task. This method, shown in Fig. \ref{fig:resshift} not only directly incorporates the input image conditions into the optimisation process but also reduces the number of diffusion steps required for sampling, thereby improving inference efficiency.

Let us denote the residual between the BF and IF images as $e_0$, i.e., $e_0 = y_0 - x_0$. Here, $y_0$ is the input BF condition, and $x_0$ is the target image. The core idea, thus, is to learn the mapping from $y_0$ to $x_0$, rather than from a pure random state $x_T$ to $x_0$, which requires long sampling iterations.

Therefore, the forward process in the modified DDPM is a transition from $x_0$ to $y_0$ by gradually shifting their residual $e_0$ through a Markov chain of length $T$. Mathematically, for each time step $t$, the transition of the image is modeled by:
\begin{equation}\label{eq:transform}
    x_t = x_{t-1} + \alpha_t e_0 + \kappa \sqrt{\alpha_t} \epsilon_t,
\end{equation}
where $\alpha_t$ represents the noise variance schedule, $\epsilon_t \sim \mathcal{N}(0, I)$, and $\kappa$ is a hyper-parameter controlling the noise variance. 

In the reverse process, a straightforward solution is described in Equation \ref{eq:reparam}, whereby the optimisation target is refined to minimise the difference between the predicted noise and the actual noise. However, this target does not directly optimise the quality of the final synthesised data, leading to unstable performance of the final synthesised image. 

Therefore, it is worth considering how to re-parameterise $\mu(x_t)$ so that we can revise the loss function from $\mathcal{L}_t = \|x_{t-1} - \mu(x_t)\|^2$ to $\mathcal{L}_t = \|x_0 - f(x_t)\|^2$, leading to direct optimization with respect to the final target IF image. 

To achieve this, we have to build up the relation between $x_{t-1}$, $x_{0}$ and $x_{t}$. According to the residual diffusion forward process in Equation \ref{eq:transform}, we have

\begin{equation}
\begin{aligned}
    x_t &= x_{t-1} + \alpha_t e_0 + \kappa \sqrt{\alpha_t} \epsilon_t \\
    &= x_{t-2} + \alpha_{t-1} e_0 + \kappa \sqrt{\alpha_{t-1}} \epsilon_{t-1} + \alpha_t e_0 + \kappa \sqrt{\alpha_t} \epsilon_t \\
    &= ....\\
    &= x_0 + \sum_{n=0}^{t} \alpha_n e_0 + \kappa \sum_{n=0}^{t} \sqrt{\alpha_n} \epsilon_n\\
    &= x_0 + \eta_t e_0 + \kappa \sqrt{\eta_t} \epsilon_t.
\end{aligned}
\end{equation}
This equation can be interpreted as using a shifting sequence $\{\eta_t\}_{t=1}^T$ to define how much residual is added at each step. Together with equation \ref{eq:transform}, we can define the relationship between $x_{t-1}$ to $x_t$ and $x_0$ as 

\begin{equation}
    x_{t-1} = \frac{\eta_{t-1}}{\eta_t} x_t + \frac{\alpha_t}{\eta_t} x_0 +  \kappa^2 \frac{\eta_{t-1}}{\eta_t} \alpha_t \epsilon_t.
\end{equation}

Thus, $\mu(x_t)$ can be re-parameterized as 
\begin{equation}
    \mu_{\theta}(x_t, y_0, t) = \frac{\eta_{t-1}}{\eta_t} x_t + \frac{\alpha_t}{\eta_t} f_{\theta}(x_t, y_0, t),
\end{equation}
resulting in the final optimization target  
\begin{equation}
\mathcal{L}_t = \frac{\alpha_t}{2 \kappa^2 \eta_t \eta_{t-1}} \| f_{\theta}(x_t, y_0, t) - x_0 \|_2^2.
\end{equation}

% \subsection{VQ-GAN and Latent Residual Diffusion}
% To reduce the time and memory costs of training, we used the latent diffusion \cite{rombach2022high} process in this paper. We operates the diffusion process in the latent space and enables the optimization of diffusion models on limited computational resources. In our paper, we mapped the images into the latent space using a pre-trained VQ-GAN  \cite{esser2021taming}. 

% VQ-GAN approximates the latent space distributions with pre-trained networks. However, finding a suitable prior distribution for continuous variables can be complicated, and the joint or conditional distributions among multiple continuous variables are difficult to derive from data-driven methods. Thus, VQ-GAN quantizes the latent features into a discrete latent space, i.e., each pixel in the latent feature maps is a $K$-way categorical variable, sampling from 0 to $K$, and by using autoregressive models that compute the conditional distributions among pixels, the latent space distribution could be approximated. 

\subsection{Instance Mask to Segmentation Boundaries}
By introducing two additional segmentation channels, our proposed model is capable of producing nuclei and cell segmentation simultaneously.

For the synthesis process, we opted not to use one-hot encoded multichannel instance segmentation masks. This decision was made to avoid the increased GPU resource demands required for training, as each image contains over 50 cell instances. Additionally, we opted not to use integer segmentation maps because their compression would result in a loss of detail for instances indexed with lower values (e.g., 1, 2, or 3). During synthesis and Gaussian noise diffusion, objects with smaller labels tend to diminish, and the residual between the integer segmentation maps also provide an unfair guidance during synthesis. 

\begin{figure}
    \centering
    \includegraphics[width=1\linewidth]{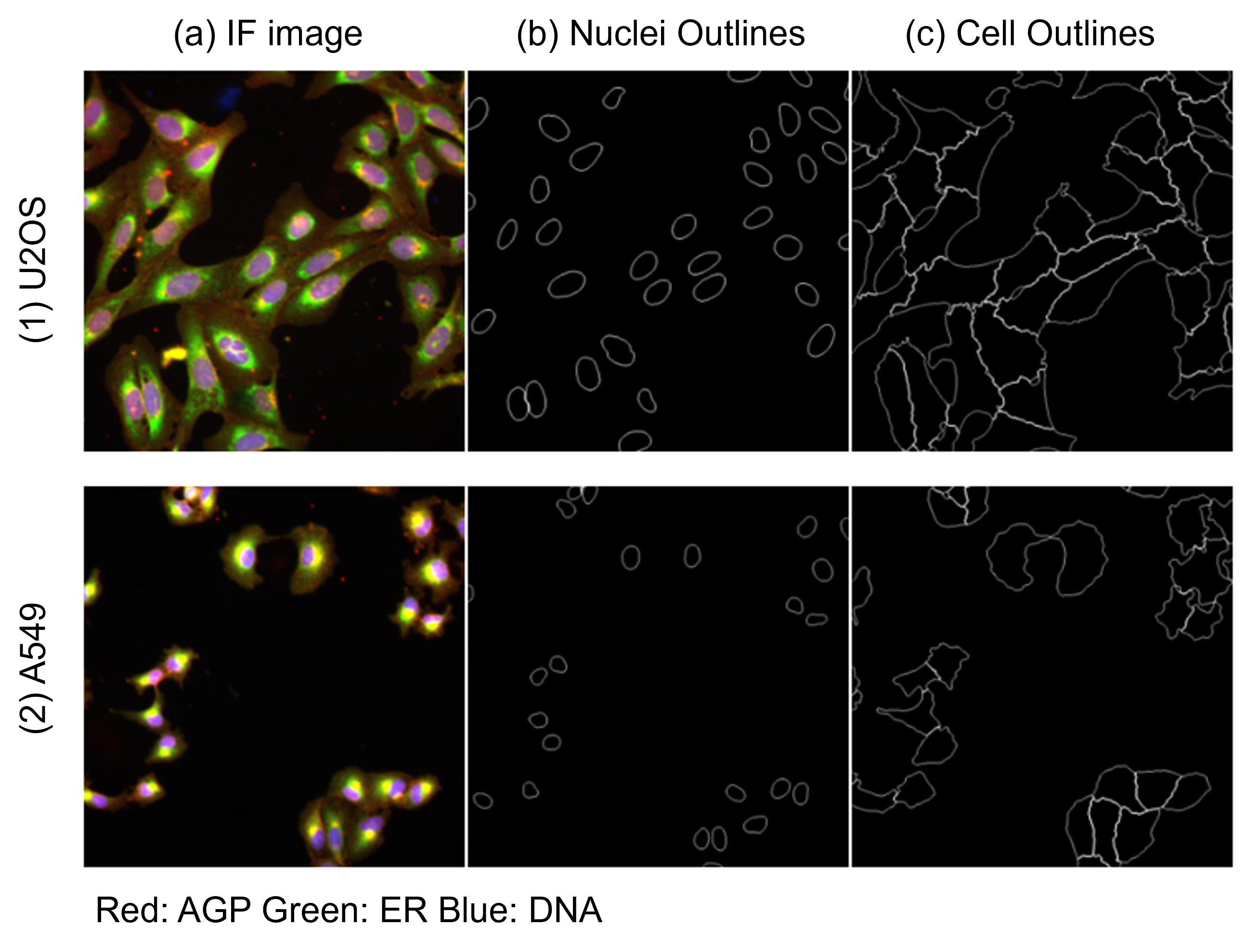}
    \caption{IF images and corresponding Nuclei (b) and Cell (c) segmentation boundaries. }
    \label{fig:example}
\end{figure}

Instead, we utilised cell boundaries for the synthesis process; example images are shown in Fig. \ref{fig:example}. The target was a binary mask to ensure the maintenance of boundary integrity and consistency. We obtained our ground truth results from CellProfiler and manually verified the boundary results.

\begin{figure*}[h]
    \centering
    \includegraphics[width=\linewidth]{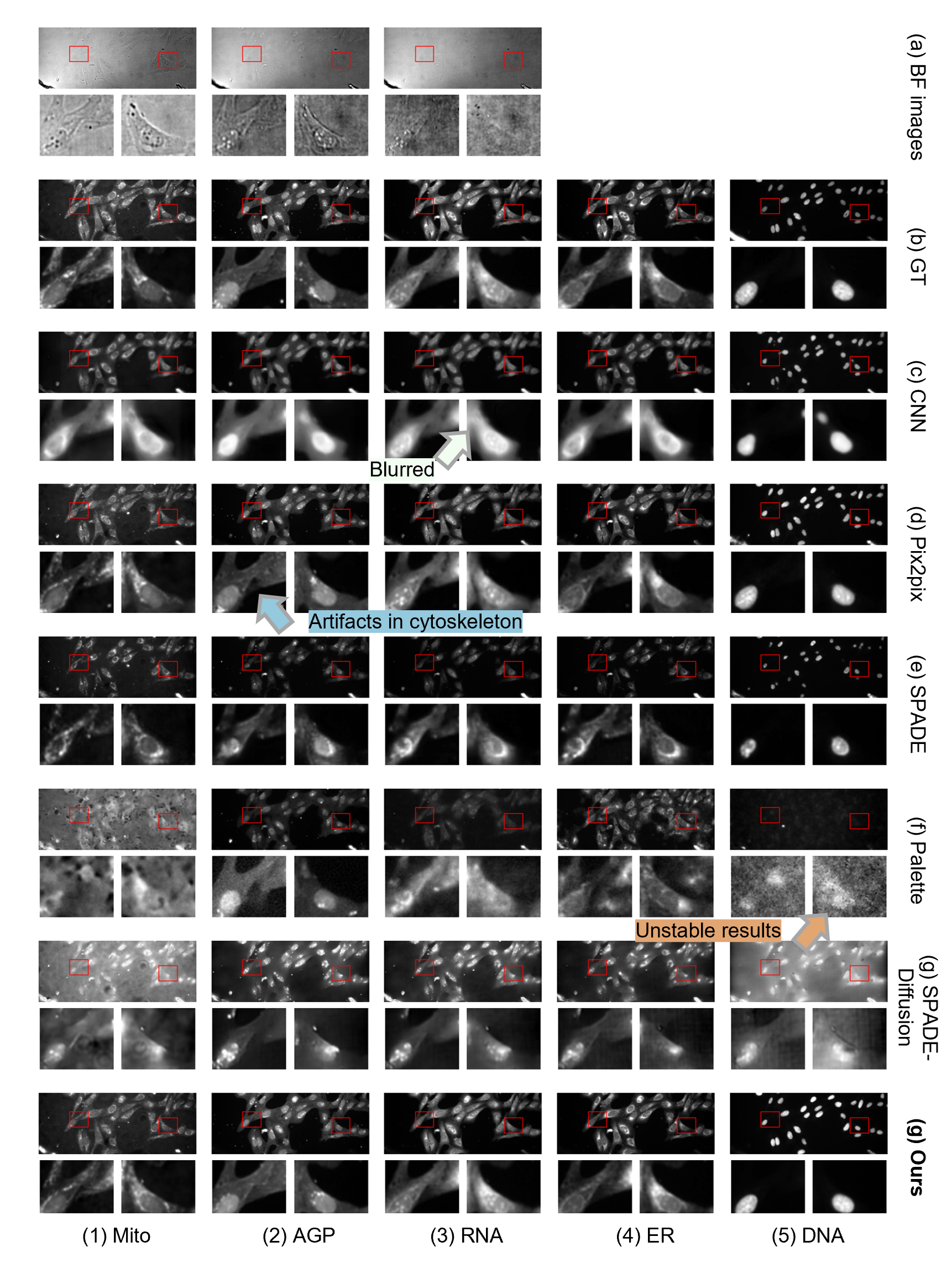}
    \caption{Visualisation of the synthesised IF images for U2OS cells demonstrates that our proposed CellResDM performs best in synthesising IF images. }
    \label{fig:vis_result}
\end{figure*}

\subsection{Evaluation Metrics}
To quantitatively assess the performance of our image-to-image synthesis models, we employed several widely recognized metrics: Mean Squared Error (MSE), Peak Signal-to-Noise Ratio (PSNR), Structural Similarity Index Measure (SSIM). Several images near the edges of the plates contained no useful information because the corresponding ground truth IF images for these regions were entirely black. To ensure more accurate analysis, we excluded these non-informative images from the dataset.

Additionally, for evaluating the accuracy of nuclei and cell segmentation, we utilised the StarDist evaluation tool\footnote{\url{https://github.com/stardist/stardist/blob/main/stardist/matching.py}}, which includes metrics such as Precision, Recall, F1 Score, and Mean True Score (the average IoU of true positives), Mean Matched Score (the average IoUs of matched true positives, providing an average score of the overlap between matched predicted and true objects), and Panoptic Quality \cite{kirillov2019panoptic}, which combines aspects of segmentation quality and recognition quality. 

Furthermore, we explored the Biological Feature Correlation to determine how well the synthesised images preserve biological details and features critical to biomedical analysis. This involves quantifying the correlation between features extracted from the synthesised images and those from the ground-truth images using methods tailored to specific biological structures and functions. 

\subsection{Methods for Comparison}
We compared the performance of our proposed method with three types of image-to-image generation models, including the UNet model \cite{lee2018deephcs}, Pix2Pix Model \cite{isola2017pix2pix}, SPADE Model \cite{park2019spade}, Palette Model \cite{saharia2022palette}, and SPADE-Diffusion \cite{wang2022spadediffusion}.

For the CNN model tasked with synthesising immunofluorescence (IF) images from bright field (BF) images, we employed the UNet architecture, which is optimised with an MSE loss \cite{lee2018deephcs}.

In the Pix2Pix model, we incorporated a VGG loss \cite{johnson2016perceptual} to enhance performance across all images. To accommodate the pretrained VGG classifier, we expanded each channel into three channels. Additionally, we evaluated the Spatially-Adaptive Denormalization (SPADE) technique for conditioning to assess its effectiveness in conditional synthesis \cite{park2019spade}.

Within the diffusion model framework, the Palette model integrates conditions in Denoising Diffusion Probabilistic Models (DDPM) with classifier-free guidance \cite{saharia2022palette}. Furthermore, we tested the SPADE technique for conditioning, employing a SPADE-Diffusion framework to explore its potential in enhancing synthesis quality \cite{wang2022spadediffusion}.

\section{Data and Experiments}
\subsection{Data and Preprocessing}
We trained our model on the public \texttt{cpg0000} dataset from the Cell Painting Gallery, which contains over three million images for both bright field and cell painting images, treated by over 300 compounds and 160 genes (CRISPR knockout) profiled in A549 and U2OS cell lines, at two time points.

\textbf{Bright Field (BF) images:} The JUMP ORF production captured 3 $Z$ positions—one equal to the lowest fluorescence position (Brightfield), one 5$\mu$m above that position (\texttt{BFHigh}), and one 5$\mu$m below that position (\texttt{BFLow}). Multiple z-planes are useful if one is interested in generating digital phase contrast images. To simplify the data for our residual models, we only use the Brightfield position (the middle focal plane). For comparison studies, we instead used all BF channels.

\textbf{Five-Channel Cell Painting images:} The channels included are nucleus (Hoechst; DNA), nucleoli and cytoplasmic RNA (SYTO 14; RNA), endoplasmic reticulum (concanavalin A; ER), Golgi and plasma membrane (wheat germ agglutinin (WGA); AGP), mitochondria (MitoTracker; Mito), and the actin cytoskeleton (phalloidin; AGP).

Overall, we selected 8 plates (4 for A549 and 4 for U2OS) for our experiments. Four plates were used for training, two for validation, and two for testing. Each plate contains 384 wells, resulting in a total of 27,648 images for all 8 channels, i.e., 3,456 image sets, with each well either treated with a specific compound or DMSO as a control. 
%[placeholder: batch effects...]

\subsection{Experimental Settings}
There are several key parameters in the residual diffusion process and in model training. We select \(\{\eta_t\}_{t=1}^T\), which corresponds to the noise scheduling in the original DDPM. To guarantee a smooth transition between \(x_0\) to \(y_0\), a coefficient \(\kappa\) is introduced to modulate the variance of the added noise in each diffusion step. For these two parameters, we choose the noise scheduling and noise variance value based on \cite{yue2024resshift}. The noise scheduling is exponential and is defined by
\begin{equation}
\sqrt{\eta_t} = \sqrt{\eta_1} \times b_0^{\beta_t}, \quad t = 2, \ldots, T-1,
\end{equation}
where
\begin{equation}
\beta_t = \left( \frac{t-1}{T-1} \right)^p \times (T-1), 
\quad b_0 = \exp \left[ \frac{1}{2(T-1)} \log \frac{\eta_T}{\eta_1} \right].
\end{equation}
We select \(p=0.3\) and \(\kappa=2\).

All models were trained and tested on NVIDIA RTX 6000 GPU clusters. To ensure a fair comparison, the batch size for training all models was set to 2. The model architecture for residual diffusion is a UNet model enhanced with Swin Transformer components. Each level of the model contains 2 residual blocks.
% Second, for the VQ-GAN model, the latent feature map size is $3\times8192$. Two residual blocks were used for each downsampling step. In our implementation, we used a pre-trained VQ-GAN model trained on natural images, optimized using the LPIPS loss and a discriminator loss. For a fair comparison, we reported two versions of CellResDM: one using a VQ-GAN trained on natural images and one without VQ-GAN. 
\begin{figure}[h]
    \centering
    \includegraphics[width=\linewidth]{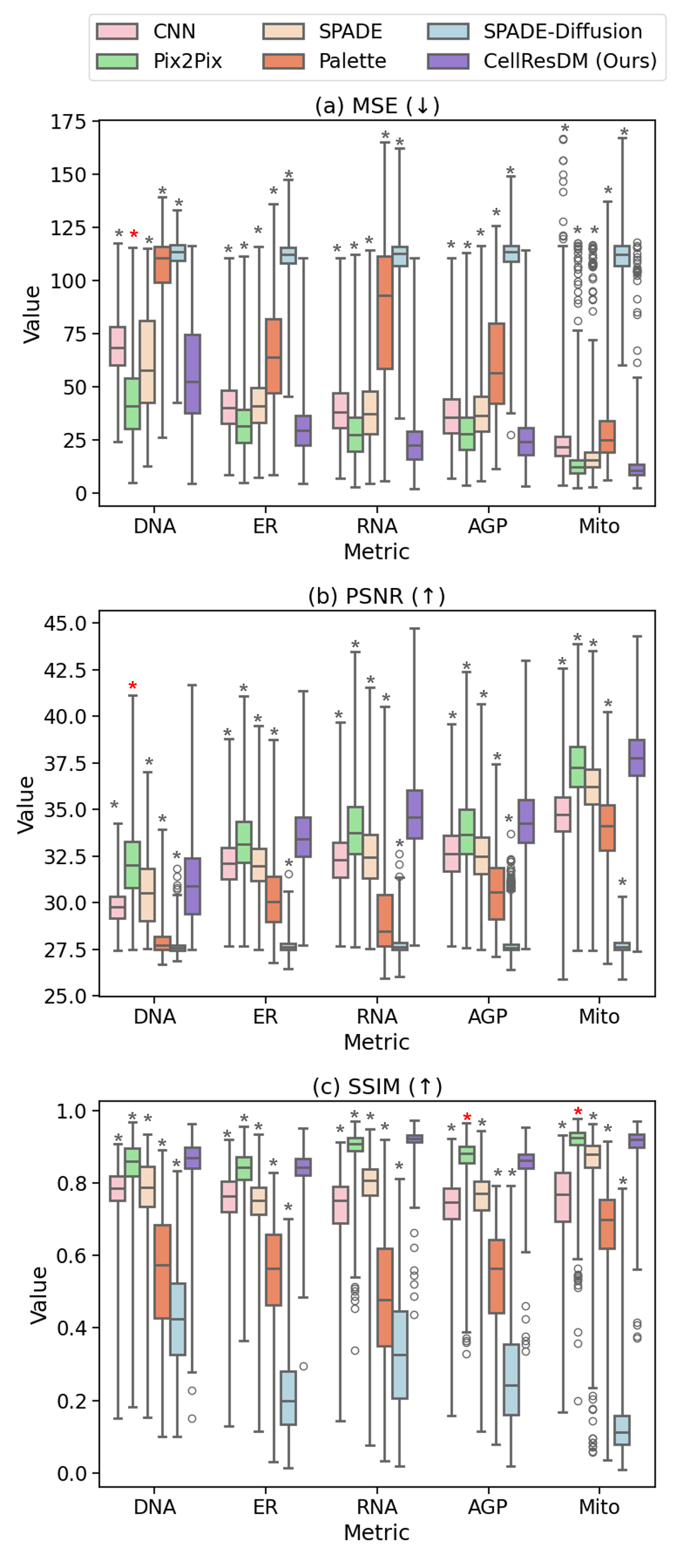}
    \caption{Quantification results of image quality metrics for various methods compared to our method. The results are grouped by their respective channels, with images normalized to a range of 0-255.}
    \label{fig:mse_result}
\end{figure}
\section{Results}
\subsection{Qualitative evaluation}
In this section, we examine and compare the visual quality of synthesised IF images from various computational models including CNNs, GANs, and diffusion models. Our proposed CellResDM demonstrated the best performance in generating IF images. 
As in Figure \ref{fig:vis_result}, the CNN method produces blur artefacts, while traditional diffusion models generated unstable image representation, which is likely the result of diffusion models depending on a de-noising process from a purely random state. It should be noted that the Pix2Pix models shows good visual appearance though the synthesised cytoskeleton is not complete. This fragmentation leads to biologically infeasible results, undermining the practical utility of the images for scientific analysis. 

It should be noted that our proposed CellResDM model only ran 15 denoising iterations, significantly reducing the sampling speed for large-scale datasets. The average sampling speed for Palette is 40 seconds per 5-channel IF image, whereas our model only takes 1.5 seconds.

\subsection{Quantitative Evaluation}
The quantitative image quality evaluation in Figure \ref{fig:mse_result} demonstrates the superiority of our proposed method in synthetic IF quality. We conducted a T-test for each channel's quality metrics between our method and all other methods, with significant differences ($p<0.05$) highlighted by asterisks (Red asterisks indicate instances where our model underperforms). To compare our CellResDM model with the second best model Pix2Pix, it  is clear from  Figure \ref{fig:mse_result} that our CellResDM model outperforms Pix2Pix in the predictions of AGP, RNA, ER, and DNA channels, and is only marginally inferior in DNA channel prediction.

\subsection{ Morphological Feature Correlation}
We calculated the correlation between hand-crafted features using CellProfiler and their mean values across different channels, as shown in Table \ref{tab:corr}. We showed that CellResDM produces features with the highest correlation with hand crafted features in each of the 8 categories for both A549 and U2OS cell lines compared to other methods. 

For a fair comparison, we used the CellProfiler pipeline to identify the location of nuclei and cell regions from the IF images, instead of using our proposed segmentation masks. It should be noted that the average sampling throughput speed for generating segmentation masks using CellProfiler is 0.03 samples per second, meaning each sample requires approximately 33 seconds to process. While this might not sound excessive, it is important to note that each plate contains more than 3000 multi-channel images. Consequently, obtaining the segmentation masks for just one plate takes more than 24 hours.

\begin{table*}[ht]
\centering
\begin{tabular}{llp{1.5cm}p{1.5cm}p{1.5cm}p{1.5cm}p{1.5cm}p{1.5cm}}
\hline
CellType & Feature Type & CNN   & Pix2Pix & SPADE & Palette & SPADE-Diffusion & Ours \\ \hline
A549     & Correlation   & 0.006 & 0.035   & -0.014 & 0.016   & -0.006           & \textbf{0.104} \\ 
         & Count         & 0.997 & 0.998   & 0.997  & 0.903   & 0.982            & \textbf{0.999} \\ 
         & Granularity   & 0.041 & 0.121   & -0.011 & 0.043   & -0.005           & \textbf{0.203} \\ 
         & Intensity     & 0.051 & 0.042   & -0.007 & -0.002  & 0.029            & \textbf{0.128} \\ 
         & Location      & 0.490 & 0.518   & 0.464  & 0.541   & 0.517            & \textbf{0.557} \\ 
         & Neighbors     & 0.353 & 0.394   & 0.282  & 0.332   & 0.291            & \textbf{0.454} \\ 
         & RadialDistribution & 0.019 & 0.023 & 0.009  & 0.005   & 0.011            & \textbf{0.098} \\ 
         & Texture       & 0.048 & 0.030   & 0.017  & 0.008   & 0.018            & \textbf{0.108} \\ \hline
U2OS     & Correlation   & 0.015 & 0.017   & -0.003 & 0.018   & 0.018            & \textbf{0.112} \\
         & Count         & 0.989 & 0.997   & 0.990  & 0.522   & 0.772            & \textbf{0.998} \\ 
         & Granularity   & 0.074 & 0.080   & 0.011  & -0.025  & 0.041            & \textbf{0.244} \\ 
         & Intensity     & 0.025 & 0.004   & 0.005  & -0.026  & -0.007           & \textbf{0.120} \\ 
         & Location      & 0.533 & 0.482   & 0.481  & 0.476   & 0.486            & \textbf{0.568} \\ 
         & Neighbors     & 0.208 & 0.355   & 0.336  & 0.351   & 0.405            & \textbf{0.438} \\ 
         & RadialDistribution & 0.033 & 0.001 & 0.001  & -0.003  & -0.006           & \textbf{0.080} \\ 
         & Texture       & 0.006 & 0.010   & 0.000  & 0.015   & 0.013            & \textbf{0.103} \\ \hline
\end{tabular}
\caption{Comparison of feature types across different methods for A549 and U2OS cell types.}
\label{tab:corr}
\end{table*}
\begin{figure}[h]
    \centering
    \includegraphics[width=\linewidth]{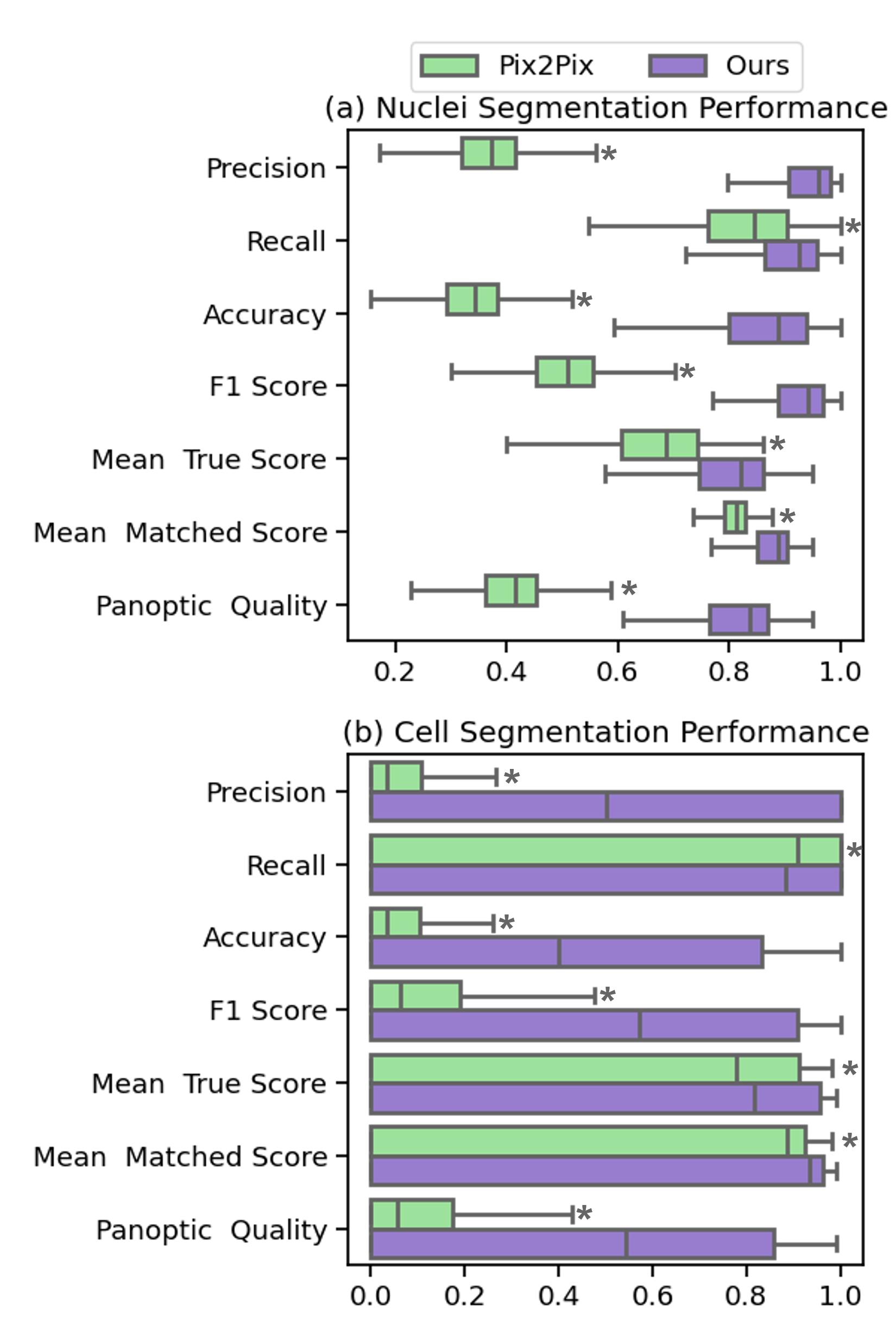}
    \caption{Segmentation performance comparison between the Pix2Pix model and our model. * indicates a significant difference as determined by a Student's t-test with $p < 0.05$}
    \label{fig:same}
\end{figure}

\subsection{Segmentation Evaluation}
\textbf{Comparison with Other IF + Segmentation Synthesis Methods:} We selected the second best-performing model, Pix2Pix, and modified it to simultaneously produce segmentation and IF images by adding two additional channels during synthesis. The segmentation performance was evaluated as detailed below. As in Fig. \ref{fig:same}, our proposed method predicts the most accurate segmentation masks, outperforming the pix2pix model which produces the second-best image quality in IF image synthesis. 

We noticed that cell segmentation is a more challenging than nuclear segmentation. This difficulty may be attributed to the fact that the cell borders, marked by WGA in the AGP channel, are not prominent in IF  images, as shown in Fig. \ref{fig:example}. However, nuclei segmentation is easier due to the clear and significant presence of boundary information in the DNA channel.

\begin{figure}[h]
    \centering
    \includegraphics[width=1\linewidth]{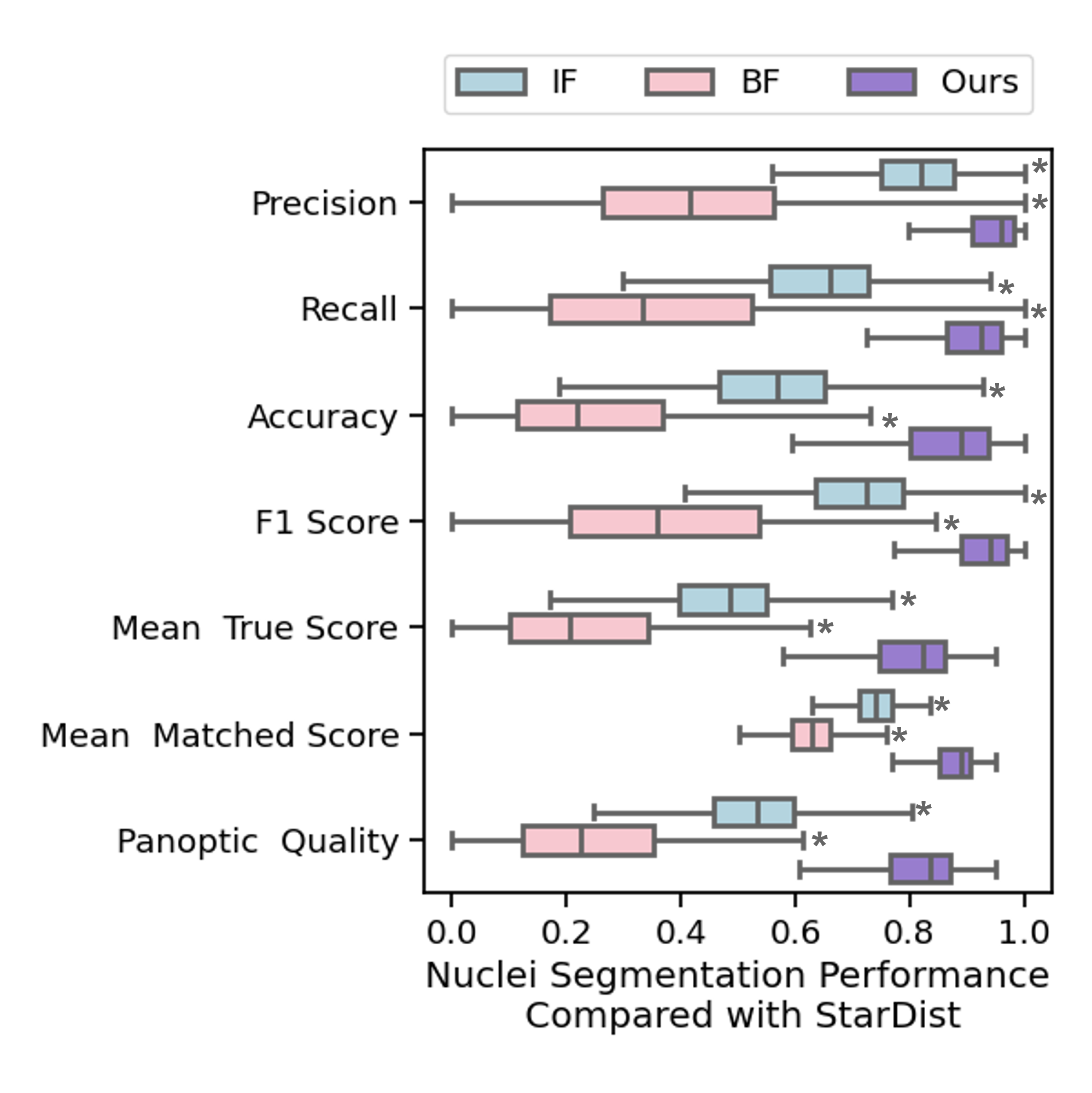}
    \caption{The nuclei segmentation result from stardist model and our algorithm. We trained two StarDist models using IF and BF images, respectively. * indicates a significant difference as determined by a Student's t-test with $p < 0.05$.}
    \label{fig:stardist}
\end{figure}
\textbf{Comparison with Popular Segmentation Algorithms:} In addition, we compared the segmentation results of nuclei and cell masks from our algorithm with segmentation models specifically trained on ground truth images and augmentations, e.g. StarDist. We trained two deep learning segmentation algorithms using StarDist \cite{stardist} on 1) BF images and 2) IF images to compare the segmentation performance on nuclei. As shown in Fig. {fig:stardist}, our simultaneous output of nuclei segmentation outperforms StarDist in all 7 quality metrics. 
%We did not compare our method with CellPose, however the comparison with StarDist clearly illustrated the good performance our proposed method.

\section{Discussion}
\subsection{Latent CellResDM: Diffusion Process in the Latent Space}

To reduce the time and memory costs of training, many researchers have projected images into a lower resolution latent space for the diffusion process \cite{rombach2022high}. In this paper, we also provide an option for latent CellResDM, facilitating IF synthesis with limited computational resources.

\begin{table}[h]
\centering
\begin{tabular}{lcc}
\hline
Model & MSE (STD) & Sampling Throughput \\
\hline
Nature VQ-GAN & 15.25 (17.57) & 0.61/s \\
Our VQ-GAN & 54.15 (56.15) & 0.53/s \\
Without VQ-GAN & 14.62 (17.19) & 0.70/s \\
\hline
\end{tabular}

\caption{Performance metrics of Latent CellResDM with different pre-trained VQ-GAN}
\label{tab:ae}
\end{table}

In this section, we map the images into the latent space using a pre-trained VQ-GAN \cite{esser2021taming} on: 1) the OpenImages dataset \cite{kuznetsova2020open}, and 2) the cell painting dataset (where we randomly selected 11 plates from the \texttt{cpg0000} dataset). We then compare this with the non-latent diffusion process to achieve a critical balance between training resources and performance. In our implementation, we use a VQ-GAN that downsamples the input resolution by a factor of 4, resulting in a latent space resolution of (512/4) x (512/4). We evaluated the image quality and GPU resources, as shown in Table \ref{tab:ae}. Since the VQ-GAN only accepts and outputs 3-channel images, we split the channels into three groups: i) Mito,AGP,DNA; ii) RNA,ER,DNA; and iii) DNA, nuclei segmentation, cell segmentation, and trained three diffusion models, respectively. We selected the DNA channel in the first model as the final output.
% \begin{figure}[h!]
%     \centering
%     \includegraphics[width=\linewidth]{figs/residual.png}
%     \caption{The absolute residual difference between the BF image and different channels of IF images.}
%     \label{fig:residual}
% \end{figure}

We noticed that the model using the pre-trained VQ-GAN trained on the cell painting dataset performs the worst in terms of image quality. This might be due to the lack of generalisability of the pre-trained VQ-GAN on the cell painting dataset and the smaller dataset size (260k images) compared to the OpenImages dataset (9.2M images) used in \cite{rombach2022high}.

\subsection{CellResDM to BF Image Artefacts}
In addition to accurately matching the true signals, we discovered that our proposed CellResDM also has the ability to faithfully model noise in microscopy images. 

\begin{figure}[ht]
\centering
\includegraphics[width=\linewidth]{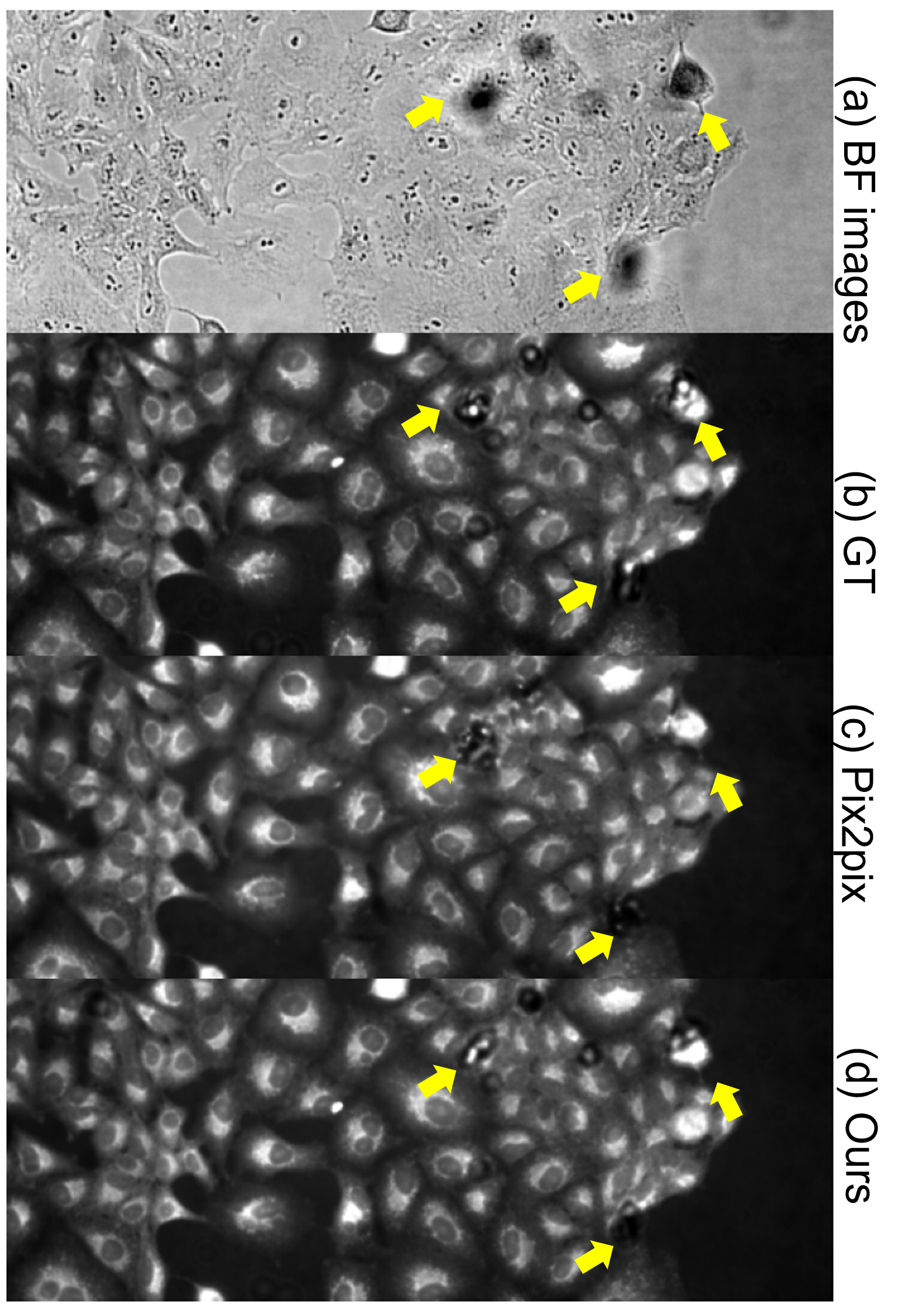}
\caption{BF images showing three out-of-focus artifacts indicated by yellow arrows (a) and the corresponding real (b) and synthetic IF (c) images of the mitochondria channel. The Pix2Pix model effectively manages the artifacts, producing realistic IF images despite the presence of imperfections in the BF images.}
\label{fig:imperfections}
\end{figure}

In Figure \ref{fig:imperfections} (a) and (b), three artifacts are evident in both the BF and mitochondria channel images, as highlighted by yellow arrows. In the BF image, these artifacts appear as black stains due to their weak signals. As a result, the pix2pix model, which depends on the BF image for input, is unable to replicate this noise.

However, our CellResDM effectively manages these artifacts, producing realistic IF images that even incorporate the noise. This demonstrates that CellResDM can accurately model the data distribution and handle imperfections in input data without the need for specific denoising or preprocessing steps.

\section{Conclusion}
Our research introduces a method called CellResDM, which employs a residual diffusion process to generate pseudo cell painting images.  In comparison with the standard DDPM model, our method generates synethetic cell painting images with the highest quality, using the shortest inference time.  Robust evaluations were performed to assess image appearances both qualitatively and quantitatively, suggesting CellResDM generates realistic images and can even model realistic noise.  In addition, our alogrithm simultaneously produces accurate segmentation masks for both nuclei and cell bodies, outperforming StarDist. In conclusion, CellResDM is a useful model for generating pseudo cell painting images and segmentation results for compound screening and profiling in drug discovery.

%% Loading bibliography style file
%\bibliographystyle{model1-num-names}
\bibliographystyle{cas-model2-names}

% Loading bibliography database
\bibliography{cas-refs}

%\vskip3pt

\end{document}